\documentclass[final]{IEEEtran}
\ifCLASSINFOpdf
\usepackage[pdftex]{graphicx}
\else
\fi
\usepackage[hyphens]{url}
\usepackage{booktabs}
 \usepackage{multirow}
 \usepackage{subfigure}

\begin{document}
\bstctlcite{IEEEexample:BSTcontrol}
\IEEEpubid{978-1-4673-2543-1/12/\$31.00 \copyright 2012 IEEE}
\title{PhishAri: \\Automatic Realtime Phishing Detection on Twitter}


\author{\IEEEauthorblockN{
\IEEEauthorblockA{Anupama Aggarwal$^\dagger$, Ashwin Rajadesingan$^*$, Ponnurangam Kumaraguru$^\dagger$  \\}
$^\dagger$Indraprastha Institute of Information Technology, India, $^*$Arizona State University, USA\\
anupamaa@iiitd.ac.in, arajades@asu.edu, pk@iiitd.ac.in}

%\and
%\IEEEauthorblockN{Author2}
%\IEEEauthorblockA{BBB\\
%Email: e2}
%\and
%\IEEEauthorblockN{Author3}
%\IEEEauthorblockA{CCC\\
%Email: e3}
}


%Telephone: (800) 555--1212, Fax: (888) 555--1212}


\maketitle

\begin{abstract}
%\boldmath
%PhishAri is a realtime system which detects phishing tweets on Twitter. PhishAri uses multiple features to detect the presence of phishing URLs in tweets. These features include - (i) Querying phishing blacklists - PhishTank and Google Safebrowsing , (ii) URL-based features - URL length, number of dots in URL, number of redirections, use of URL shorteners etc., (iii) Tweet-based features - length of tweet, use of hashtags, retweet count etc, (iv) Source-based features - Follower/Followee ratio of the user who posted the tweet, number of tweets and age of the source user account etc.,(v) WHOIS-based features - ownership period of domain name, registrar through which domain was bought etc. The combination of these features helps detect phishing on Twitter and an easy-to-use browser plugin keeps Twitter users safe from phishing scams. The plugin provides a clean and intuitive interface to indicate phishing tweets. It detects phishing links in realtime, thus warning users from clicking malicious URLs. PhishAri browser plugin displays a legitimate or phish indicator (next to the URL) for tweets containing URLs. Phishing URLs are marked with a red indicator, warning users not to click on the same while the legitimate links, on the other hand, are marked with a green indicator.
%Phishing is a common method of identity theft by fraudulently acquiring personal and confidential credentials. Phishing scams have been traditionally carried out by sending impersonating emails. However, 

With the advent of online social media, phishers have started using social networks like Twitter, Facebook, and  Foursquare to spread phishing scams. Twitter is an immensely popular micro-blogging network where people post short messages of 140 characters called tweets. It has over 100 million active users who post about 200 million tweets everyday. Phishers have started using Twitter as a medium to spread phishing because of this vast information dissemination. Further, it is difficult to detect phishing on Twitter unlike emails because of the quick spread of phishing links in the network, short size of the content, and use of URL obfuscation to shorten the URL. Our technique, \emph{PhishAri}, detects phishing on Twitter in realtime. We use Twitter specific features along with URL features to detect whether a tweet posted with a URL is phishing or not. Some of the Twitter specific features we use are tweet content and its characteristics like length, hashtags, and mentions. Other Twitter features used are the characteristics of the Twitter user posting the tweet such as age of the account, number of tweets, and the follower-followee ratio. These twitter specific features coupled with URL based features prove to be a strong mechanism to detect phishing tweets. We use machine learning classification techniques and detect phishing tweets with an accuracy of 92.52\%. We have deployed our system for end-users by providing an easy to use Chrome browser extension. The extension works in realtime and classifies a tweet as phishing or safe. In this research, we show that we are able to detect phishing tweets at zero hour with high accuracy which is much faster than public blacklists and as well as Twitter's own defense mechanism to detect malicious content. We also performed a quick user evaluation of PhishAri in a laboratory study to evaluate the usability and effectiveness of PhishAri and showed that users like and find it convenient to use PhishAri in real-world. To the best of our knowledge, this is the first realtime, comprehensive and usable system to detect phishing on Twitter.
\end{abstract}


% \fi
\IEEEpeerreviewmaketitle

\section{Introduction 
%/ Research Motivation and Aim
}\label{section:introduction}

\begin{figure}[!ht]
\centering
\includegraphics[scale=0.3]{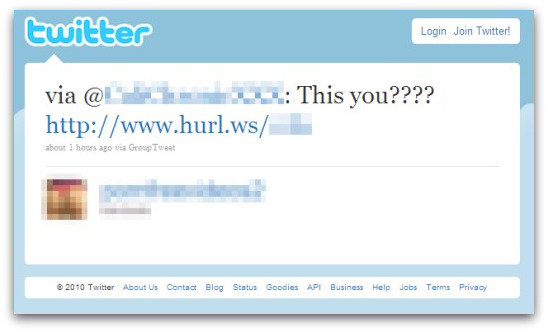}
\caption{An example of a phishing tweet. The URL which appears in the tweet redirects the user to a fake Twitter login page.}
\label{fig:phishing_twitter}
\end{figure}

%\subsection{Rise of Phishing on Online Social Media}
Phishing is an online fraudulent technique used to acquire personal and confidential credentials. Phishing attacks lead to theft of sensitive information such as e-commerce accounts, confidential bank account details and other personally identifiable information of an Internet user. Such attacks have disastrous consequences as they result in identity theft and often result in huge monetary loss~\cite{jakobssonphishingcountermeasures:2006}.
%A report by RSA, the Security Division of EMC 
It is estimated that \$520 million were lost worldwide from phishing attacks in 2011 alone.~\footnote{\url{http://www.rsa.com/solutions/consumer_authentication/intelreport/11541_Online_Fraud_report_1011.pdf}} Traditionally, phishing attacks target email users, however, with the unprecedented explosion in popularity of Online Social Media (OSM) like Facebook, Twitter, YouTube and Foursquare, adversaries also use these media to spam and phish. In 2010, 43\% of all the OSM users were targets of phishing attacks.~\footnote{\url{http://www.infographicsarchive.com/social-media/the-dark-side-of-social-media-how-phishing-hooks-users/}} In 2012, around 20\% of all phishing attacks targeted Facebook.~\footnote{\url{http://www.securelist.com/en/analysis/204792234/Spam_report_May_2012}} Another report in 2012 suggests that social network phishing has jumped 221\% to 9,974 attacks during Q1 of 2012 when compared to such phishing instances in the previous quarter.~\footnote{\url{https://www.markmonitor.com/mmblog/q1-2012-fraud-intelligence-report/}} There has been an increase in phishing attacks through social media due to ease and spread of information on social networks. Multiple instances of phishing attacks have been reported on Facebook~\footnote{\url{http://www.barracudalabs.com/wordpress/index.php/2012/04/06/warning-new-facebook-phishing-via-facebook-chat-and-note/}}, Twitter~\footnote{\url{http://mashable.com/2011/10/26/warning-twitter-spam/}} and  other OSMs~\cite{chhabra2011phi}. Such a rise in phishing attacks on social media presents a dire need for technological solutions to deter these attacks and protect users from phishing scams. Detecting  phishing on social media is a challenge because of (i) large volume of data -- social media allow users to easily share their opinions and interests which results into large volumes of data and hence, make it difficult to mine and analyze; (ii) limited space -- social media often impose character limitation (such as Twitter's 140 character limit) on the content due to which users use shorthand notations. Such shorthand notation is difficult to parse since the text is usually not well-formed; (iii) fast change -- content on social media changes very rapidly making phishing detection difficult; and (iv) Shortened URLs -- researchers have observed that more than half of the phishing URLs are shortened to obfuscate the target URL and to hide malignant intentions rather than to gain character space~\cite{chhabra2011phi}. Short URLs not only hide the target URL but also help in evading blacklists.
\IEEEpubidadjcol
%\subsection{Phishing on Twitter}
Twitter is an online social networking website which allows its users to, among other things, micro-blog their daily activity and talk about their interests by posting short 140 character messages called tweets. Twitter is immensely popular with more than 100 million active users who post about 200 million tweets everyday.~\footnote{\url{http://blog.twitter.com/2011/09/one-hundred-million-voices.html}} Ease of information dissemination on Twitter and a large audience, makes it a popular medium to spread external content like articles, videos, and photographs by embedding URLs in tweets. However, these URLs may link to low quality content like malware, phishing websites or spam websites. Recent statistics show that on an average, 8\% tweets contain spam and other malicious content~\cite{grier2010spam}. Figure~\ref{fig:phishing_twitter} shows an example of a malicious phishing tweet. 

%\subsection{PhishAri}
In our research, we propose PhishAri~\footnote{`Ari' in Sanskrit means Enemy, since we were building a tool to curb phishing, our system is christened PhishAri.} -- a tool to automatically detect phishing tweets in realtime. PhishAri uses various features such as the properties of the suspicious URL, content of the tweet, attributes of the Twitter user posting the tweet and details about the phishing domains to effectively detect phishing tweets. PhishAri decides whether a tweet is ``phishing" or ``safe" by employing machine learning techniques using a combination of the aforementioned features. Also, we have built a Chrome browser extension to provide realtime phishing detection to Twitter users. The browser extension protects the user from falling prey to phishing attacks by appending a red indicator to phishing tweets. Further, PhishAri is time efficient, taking an average of only 0.425 (more details later in the paper) seconds to detect phishing tweets with high accuracy of 92.52\%. Such low computation times make it ideal for real world use.

%{\bf PK:  is this really small though? probably yes, but i just hope reviewers dont beat it saying, it should be in milli seconds or something like that AR: The time also depends on the internet speeds available to the end user right? the actual classification takes less than 0.1 seconds according to my calc. I have attached a screenshot}

%Safe tweets are marked by a green indicator and phishing tweets by a red indicator. 

Our major contributions of this research work are: 
\begin{itemize}
\item 
%{\bf AR: The WARNING BIRD paper also describes a realtime phishing detection system, so I'm not sure if we can claim PhishAri to be the first such system, any thoughts?} 
Automatic realtime phishing detection mechanism for Twitter: There have been studies on phishing detection in emails and spam detection on Twitter, but, to the best of our knowledge, this is the first comprehensive focused study on realtime detection (with a focus on building usable system) of phishing on Twitter. 
\item More efficient than plain blacklisting method: Our technique proves to be better than plain blacklist lookup which is the most common technique used for phishing detection. 
%We use URL and Twitter specific features to detect phishing tweets.
\item Better than Twitter's own phishing detection mechanism: Twitter has its own phishing and malware detection mechanism but it is often thwarted by the use of URL shorteners and multiple redirections. PhishAri is able to detect more phishing tweets than Twitter's own detection mechanism. 

%{\bf PK: this is cool, but we need strong numbers to prove this. AA:will add in result section}
%\item Use of URL + Twitter specific features for phishing detection: Instead of just using the conventional URL analysis of the suspicious link, we analyze the Twitter specific properties of the suspicious tweet.
\item Real-world implementation of the system: To the best of our knowledge, PhishAri browser extension and API are the first ever deployed systems for phishing detection which can be (are being) used by real world Twitter users. PhishAri browser extension is freely available on Chrome Web Store for download.~\footnote{\url{https://chrome.google.com/webstore/detail/pheokmlohglcpigbnbenbimcombeoolm}} 
\end{itemize}

%After a thorough comparison of our technique with previous closely related work in Section~\ref{section:related}, we describe the details of our methodology. 

In this study, since our goal was to detect Phishing on Twitter and also build end-user solution for Twitter users which works in realtime, we divide our study into two parts. In the first part, we collected true positive data (described in Section~\ref{section:data}) and identified features (described in Section~\ref{section:features}) which can be used to detect phishing tweets. Based on these features we used various machine learning classification techniques to classify tweets as phishing or safe. More details of the classification experiment are described in Section~\ref{section:experiment}. We evaluated the performance of various machine learning classification methods and found the classification algorithm which works best for phishing detection on Twitter. We present these detailed results in Section~\ref{section:results}. In the second part of the study, we used the results from the first part to create a realtime usable system. We built a supporting HTTP POST API and a user-friendly Chrome extension (explained in Section ~\ref{section:phishari_api}) to detect phishing on Twitter. The API and the Chrome extension enable Twitter users to use our system and get notification about the status of a tweet as `phishing' or `safe' in realtime. With the help of a lab study described in Section~\ref{section:phishari_extension}, we also show that Twitter users find it convenient to use PhishAri Chrome extension.

We describe some of the most related work on detection of phishing in Section~\ref{section:related}. We used results, and other observations to have an in-depth discussion which is described in Section~\ref{section:discussion}, followed by some suggested future work in Section~\ref{section:future}. We end the paper with a conclusive summary of the work described in Section~\ref{section:conclusion}.

%The rest of the paper is arranged as follows: Section~\ref{section:related} describes related work done in this field. We describe the technique used by PhishAri for phishing detection, by first describing the data collection in Section~\ref{section:data} followed by the features used for phishing detection in Section~\ref{section:features}. Section~\ref{section:phishari_api} explains how we built a realtime user-friendly browser extension for Twitter. Section~\ref{section:experiment} describes the experimental setup and Section~\ref{section:results} elaborates the results of this study. We discuss the conclusion and future work in Section~\ref{section:future}. 

% followed by conclusion in Section~\ref{section:conclusion} 

%{\bf PK:  do we need future and conclusion in two different sections? lets wait for how the content folds and then we decide what to do with the sections?}.

\section{Related Work}\label{section:related}
Phishing is an online fraudulent technique to acquire personal and confidential credentials of Internet users~\cite{jakobssonphishingcountermeasures:2006}. Adversaries use phishing for various malicious activities like stealing login credentials of bank accounts, e-commerce accounts and other sensitive information of an Internet user. This section gives an overview of studies which describe how and why phishing attacks are successful and techniques used to detect phishing scams.

\subsection{Detection of Phishing Emails and Websites}
Traditionally, phishing attacks target email users. Usually,  such emails are sent through fake SMTP messages~\cite{chandrasekaran2006phishing} or by impersonating the sending authority~\cite{moore:an-empirical-analysis-of-:2007:yuqfj, moore2007examining}. There are powerful email spam filters which effectively filter out spam and phishing emails~\cite{chandrasekaran2006phishing, fette2007learning}. Fette et al. used machine learning technique to classify an email as phishing or not by using features such as age of URL, number of dots in URL and HTML content of email while obtaining a high accuracy of 99.5\%~\cite{fette2007learning}. 

Other techniques have also been extensively used to detect phishing websites. Justin et al. use lexical and host-based features of the URLs to detect malicious webpages. However, since spammers keep changing their attacking strategy, only the URL features can be difficult to detect malicious URLs~\cite{ma2009identifying}. Zhang et al. proposed CANTINA, an approach to detect phishing websites by examining the content of the website. CANTINA tries to find out whether the website has been indexed by popular search engines (e.g. Google) or not, which is considered as a measure of a legitimate website~\cite{zhang2007cantina}. CANTINA  analyzes the content of the website, identifying among other heuristics, the top five terms with highest tf-idf which are then used to determine if the website is phishing or not by feeding them to a search engine. Whereas, Phishari uses url, tweet, WHOIS and twitter based features (and does not analyze the content of the website) in making making the classification decision. CANTINA+ is another technique proposed by Xiang et al. which extracts features of a website like URL properties, webpage properties and then uses machine learning technique to classify the websites as phishing or legitimate~\cite{xiang2011cantina+}. Blacklist is another popular method in which a record of phishing websites on the Internet is maintained. These blacklists (like APWG blacklist and Google Safebrowsing) are used by many web-based toolbars and web browsers as an early warning mechanism to stop users from visiting the malicious websites. However, blacklisting technique is ineffective as most blacklists catch less than 20\% phishing websites at zero-hour~\cite{sheng2009empirical}. Other methodologies to deter phishing by spreading awareness amongst Internet users have also been developed, which include games~\cite{sheng:anti-phishing-phil:-the-d:2007:yuqfj} and educational technologies~\cite{kumaraguru2007protecting}. 

%For a detailed analysis of countermeasures for phishing, please visit~\cite{kumaraguru:phishguru:-a-system-for-e:2009:rcrwd}. 

\subsection{Phishing and spamming on Online Social Media}
With the unprecedented explosion in popularity of Online Social Media (OSM) like Facebook~\cite{gao2010detecting}, Twitter~\cite{lee2011seven} and Youtube~\cite{benevenuto2009detecting}, adversaries have started using these media to spread spam and phishing scams. In 2010, 1\% of the total Facebook users have been victims of phishing attacks, which amounts to 5 million Facebook users.~\footnote{\url{www.antiphishing.org/reports/apwg_report_Q1_2010.pdf}} Further, Twitter receives a high spam URL clickthrough rate of 0.13\%, which is much more than that of email spam~\cite{grier2010spam} as spammers take advantage of the trust network of the social media user. The ease of sharing information on OSM and the larger reach to Internet users makes it a vulnerable target to spread scams~\cite{jagaticsocialphishing:2006}.  

Spam detection studies on Twitter usually involve machine learning classification techniques. These studies highlight important twitter specific features used for spam detection, such as follower-followee ratio, tweet count and age of account. These features can be used to detect spam tweets~\cite{ wang2010don, chu2010tweeting} and spammer~\cite{benevenuto2009detecting} with high accuracy. The use of URL shorteners on Twitter to share links makes automatic detection an even more arduous task~\cite{chhabra2011phi, antoniades2011we}. There have also been studies to understand the social network of criminals and spammers on Twitter. Chao et al. found that criminal accounts are socially connected and form a small closed network~\cite{Yang:2012:ASS:2187836.2187847}. However, very little research work has been done on phishing detection on Twitter or other OSMs, and in particular, on realtime detection. 

\subsection{Real Time Detection}

Phishing is a harmful form of spam. Phishing attacks not only cause the leakage of personal information but also results in huge monetary loss. Hence it is important to build effective realtime phishing detection mechanisms for every OSM to protect its users. There exist browser based toolbars to detect phishing websites~\cite{Zhang07phindingphish}, but these toolbars require the user to click on suspected and possibly malicious URL. Thomas et al. proposed Monarch, a realtime malware and phishing detection system which crawls URLs submitted to a web service and assesses them in realtime to classify them as spam or legitimate~\cite{thomas2011design}. Monarch relies on features of the landing page which sometime may not be available. However, these solutions are not specific to Twitter. We believe that phishing detection in Twitter hosts a wide range of challenges specific to Twitter itself such as quick spread of information and the limitation of 140 characters in tweets. A dedicated solution proposed exclusively for Twitter by Lee et al. is the WarningBird system which does not focus on detecting phishing but on suspicious URLs in general~\cite{leewarningbird}. It uses correlated redirect chains of URLs on Twitter to detect phishing URLs. However, WarningBird may fail if the spammers use short redirect chain or multiple page-level redirects. Though WarningBird finds suspicious URLs on Twitter in realtime, unlike PhishAri, it does not provide an end-user mechanism for users to use and protect themselves from malicious URLs.

\subsection{Real Time Phishing Detection on Twitter}
After reviewing the above techniques, it was evident that there was very little work done to detect phishing on Twitter in realtime. To fill this gap, we designed and developed PhishAri; it leverages the power of blacklisting as well as other Twitter based, URL based and WHOIS based features. Apart from a robust API which performs realtime phishing detection, we also developed a browser-based extension to protect users from phishing attacks.

\section{Data Collection and Labeled Dataset}\label{section:data}
\begin{figure}[ht]
\centering
\includegraphics[scale=0.5]{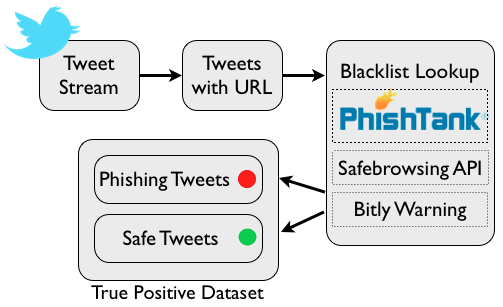}
\caption{Architecture for data collection. We collected tweets with URLs from Twitter stream and compared the URLs against phishing blacklists to build a true positive dataset.}
\label{fig:data-collection}
\end{figure}

In this section, we describe how we collected data for analysis and to build a true positive dataset of phishing tweets containing phishing URLs for our study. Data collection involves two steps as shown in Figure~\ref{fig:data-collection}, (i) collecting data from Twitter, (ii) labeling the tweets as phishing or legitimate.~\footnote{We use `legitimate' and `safe' interchangeably.}

%{\bf PK: For Architecture Figure: Add one or two more lines in the in the caption; you want to start from Twitter + API before the stream, use bird; and phishing and safe put the red and green circles. AA: Now??}

\subsection{Crawling Twitter}
For our study, we required only tweets containing URLs. We used the Twitter Streaming API~\footnote{\url{https://dev.twitter.com/docs/streaming-apis/streams/public}} and the ``Filter" function provided by the API to collect such tweets. As the Twitter Streaming API is rate limited, we can collect only a limited number of tweets per hour. In total, we collected 309,321 such tweets from 1 February 2012 to 19 April 2012.

\subsection{Labeling Tweets as Phishing or Legitimate}
To initially label tweets as phishing or legitimate in order to create an annotated dataset, we used two blacklists, PhishTank and Google Safebrowsing. For URL in every tweet, we queried both PhishTank and Google Safebrowsing APIs. 
PhishTank~\footnote{\url{http://www.phishtank.com/}} is a public crowdsourced database of phishing URLs. The suspicious URLs are submitted in the PhishTank database by contributors and marked as phishing or legitimate by volunteers. The PhishTank API accepts an HTTP POST request along with the query URL, and returns a JSON object in response which tells whether the query URL is phishing or not. Google Safebrowsing~\footnote{\url{https://developers.google.com/safe-browsing/}} is a database of malware and phishing links maintained by Google Inc. The Google Safebrowsing API uses an HTTP POST request to query the URL and matches the hash of the URL in its database of phishing and malware URL hashes. The response from the API is a JSON string describing whether the URL is ``phishing," ``malware" or ``safe." In case the URL in a tweet is phishing according to PhishTank or Google Safebrowsing API, we mark the Tweet as ``phishing." However, the inherent problem of blacklists is that they are slow to capture malicious URLs~\cite{sheng2009empirical}. We observed that the phishing URLs did not get caught by blacklists on the same day they were  posted on Twitter. Even after one day very small number of URLs were detected as phishing. Therefore, we waited for 3 days and checked all the URLs in the tweets we had collected 3 days earlier. We then repeated the same process for entire period of the data collection to build the true positive dataset of phishing tweets.
%Apart from this, we also use the warning indications by Twitter and Bitly to mark a tweet ``phishing".
%\subsubsection{Twitter and Bitly Warning}

Apart from using PhishTank and Google Safebrowsing API, we also mark tweets as ``phishing" which are declared `phishing' by Twitter itself. Twitter opens a warning page when one clicks a malicious URL. Also, many URLs posted on Twitter are shortened using Bitly URL shortening service and have the domain name ``http://bit.ly/." Bitly uses blacklisting services from various resources and also throws a warning page if it detects a phishing URL. We mark any such URL as ``phishing". Those tweets which do not have any phishing URL using this technique are marked as ``safe". After applying the above technique to 309,321 tweets, we obtained 1,589 phishing tweets (with 903 unique URLs) in our labeled dataset.

\section{Feature Selection for Phishing Detection}\label{section:features}
Phishing detection on emails has been studied in the past which shows that phishing websites can be detected using a thorough analysis of the URL and the website content. However, it has been observed that phishers constantly keep changing the techniques they use for phishing, making detection more difficult. Therefore, in this study we combine a variety of features to provide a more robust, water-tight and efficient detection methodology. This section explains the various features we identify for phishing detection on Twitter. Table~\ref{tab:features_table} gives a list of all features which we used for our analysis.

\begin{table*}[htbp]
  \centering
  \caption{Features used in PhishAri. Classified into URL based, Tweet based, Network based, and WHOIS based.}
    \begin{tabular}{p{2.1cm}p{3.5cm}l}
    \toprule
    \multirow{6}[12]{*}{URL Based (F1)} & Length of URL & Length of expanded URL in number of characters \\
          & Number of dots & Number of dots ( . ) used \\
          & Number of subdomains & Number of subdomains (marked by /) in the expanded URL \\
          & Number of Redirections & Number of hops between the posted URL and the Landing page \\
          & Levenshtein distance between redirected hops & Avg Levenshtein distance between length of redirected URLs between original \& final URL \\
          & Presence of conditional redirects & Whether the URL is redirected to different landing page for browser or an automated program \\
        \midrule
        
           \multirow{3}[6]{*}{WHOIs Based (F2)} & Registering domain name & Name of the domain provider \\
          & Ownership period & Age of the domain \\
          & Time taken to create Twitter account & How much time lapsed between creation of domain and the Twitter account \\
          
         \midrule

    \multirow{6}[12]{*}{Tweet Based (F3)} & Number of \#tags & Number of topics mentioned in tweet \\
          & Number of @tags & Number of Twitter users mentioned in tweet \\
          & Presence of trending \#tags & Number of topics mentioned which were trending at that time \\
          & Number of RTs & Number of times the tweet was reposted \\
          & Length of Tweet & Length of tweet in number of characters \\
          & Position of \#tags & Number of characters of tweets after which the \#tag appears \\
        \midrule

    \multirow{7}[12]{*}{Network Based (F4)} & Number of Followers & Number of Twitter users who follow this Twitter user \\
          & Number of Followees & Number of Twitter users who are being followed by this Twitter user \\
          & Ratio of Followers-Followees & Number of Followers / Number of Followees \\
          & Part of Lists & Whether the Twitter user is part of a public list \\
          & Age of account & How old the Twitter account is \\
          & Presence of description & Whether the Twitter account has a profile description \\
          & Number of Tweets & Number of tweets posted by the Twitter user \\
 
    \bottomrule
    \end{tabular}%
  \label{tab:features_table}%
\end{table*}%

\subsection{URL based Features}
URL features are based on the analysis of the URL of the suspicious website. The length of the URL, number of dots and subdomains, and the length of the domain are some of the most significant features that aid in phishing detection. In phishing websites, the length of the URL tends to be much longer than legitimate websites. However, the phishing domains (without TLD portion) are shorter than the regular domains. Also, phishing URLs often contain more number of dots and subdomains than legitimate URLs~\cite{fette2007learning}. %\paragraph{URL redirection}
We also observe that many phishing URLs (using ``robots.txt") automatically redirect bots (not browsers) to a legitimate domain instead of redirecting to the original phishing domain. This is one of the most effective techniques used by phishers to evade bot-based automatic detection systems. We add such behaviour also as a feature in phishing detection. We also use number of redirections as one of the features since malicious URLs often have multiple URL redirects to escape detection by blacklists.

\subsection{WHOIS based Features}
WHOIS is a query and response protocol which provides information such as ownership details, dates of domain creation / updation of the queried URL. We can identify tweets containing phishing URLs by identifying WHOIS based features that are common to phishing links. Most phishing campaigns register domains of websites from the same registrar, hence tracking the registrar may aid in detecting phishing. Further, most phishing urls are bought for a short period of one year as  offenders need to keep constantly changing the url domain names to evade blacklists. Also, the phishing domains are usually created / updated just before they are tweeted. Thus, phishing links generally have low time interval between the domain creation / updation date and the tweet creation date. Therefore, we use WHOIS based features such as registrar's name, ownership period, time interval between domain creation / updation and tweet creation date to further enhance our phishing detection methodology.

\subsection{Tweet based Features}

Malicious tweets are often tailored to gain more visibility in Twittersphere. Phishers achieve high visibility by carefully using tags in their tweets and by timing their tweets at appropriate intervals of time. Twitter provides two kinds of tags:
\begin{itemize}
\item Hashtags (\#): Indicates a topic on Twitter. An example of hashtag is \#Euro2012 which signifies the Euro Cup held in 2012. Users who post tweets about Euro Cup append \#Euro2012 in the text of their tweet. 
\item Mention tags (@): The @tag is used to either mention a fellow Twitter user or reply to one of his tweets. The tweets with @tags are displayed in the mentioned user's timeline. For example, a tweet with @John will appear in John's profile where `John' is a Twitter username.
\end{itemize}

%Figure~\ref{fig:tweet_euro} shows an example of a tweet with a hashtag and an @tag. 

%{\bf PK:  is this figure really needed??? if we are running short of space, this will the first thing to cut.... }

%\begin{figure}[!ht]
%\centering
%\includegraphics[scale=0.5]{tweet_euro.png}
%\caption{A tweet with a trending hashtag \#Euro2012 and an @tag mention to another Twitter user.}
%\label{fig:tweet_euro}
%\end{figure}

%{\bf AR: Rephrased info on hash tag and added info on @tags}

Twitter facilitates searching tweets based on topics. One who is interested in the Euro Cup can search for \#Euro2012 to obtain a list of Euro Cup related tweets posted on Twitter. When the topics are very popular, the hashtag or topic become a ``trending" topic. Trending topics are always displayed on a user's Twitter homepage (depending on their settings for the location). Thus, malicious users hijack such trending topics by posting phishing tweets with popular trending hashtags irrespective of their relevance to increase their reach and visibility. Also, the @tag allows any user to direct tweets to any other user in Twittersphere irrespective of whether they are friends / followers. Malicious users take advantage of this feature and direct phishing tweets to random users through the @ tag. Thus, malicious tweets have higher number of hashtags and @tags so that the tweet is directly visible to the mentioned users and the users searching for a topic on Twitter using the mentioned hashtags. Hence, we include such tweet based features for phishing detection.

\subsection{User Attributes and Network based Features}

Friend relationships on Twitter are unidirectional and described by the following: 
\begin{itemize}
\item Followers of a user X are those Twitter users who subscribe to X's tweets. Whenever X posts a tweet, it appears in his follower's timeline 
\item Followees of a user X are those Twitter users whom X has subscribed to. X gets all the tweets posted by his followees in his timeline.
\end{itemize} 
Studies on Twitter spam show that spammers have different tweeting behavior when compared to legitimate users. For example, spammers often post automated tweets in large numbers usually at predefined intervals of time ~\cite{benevenuto2010detecting}. Also, it has been observed that malicious users have a large number of ``followees" but a small number of ``followers." Thus, we use features such as number of tweets posted, Follower-Followee ratio and other Twitter profile information like the description of the Twitter user and presence of profile image for phishing detection.  
%Malicious users often follow a large number of Twitter users but a very few users follow them back. Due to automated activity, many times, these accounts are blocked by Twitter. Hence, We use this user behavior of the Twitter user posting a tweet for phishing detection.

\section{PhishAri API and Browser Extension}\label{section:phishari_api}
Our goal in this research work is to provide realtime protection from phishing to Twitter users. To enable this, we built a browser extension for Twitter and a supporting API to indicate whether a tweet is phishing or not.

\subsection{Browser Extension}
\begin{figure}[!h]
\centering
\includegraphics[scale=0.26]{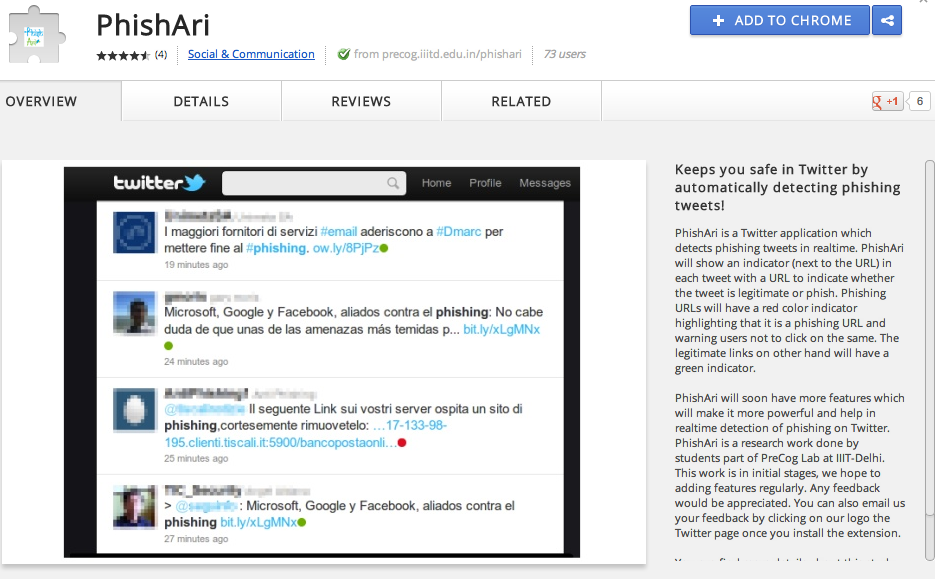}
\caption{PhishAri on Chrome. Currently, there are more than 70 active users using the extension. The green indicator shows that the tweet is `safe' whereas, a red indicator appears in front of `phishing' tweets. }
\label{fig:extension}
\end{figure}

A large fraction of Twitter users use web browser to access Twitter.~\footnote{\url{http://blog.twitter.com/2010/09/evolving-ecosystem.html}} Users are usually hesitant to change the platforms they use. Therefore, we built a browser extension which seamlessly integrates phishing detection results into the user's Twitter pages. The extension once installed shows a green indicator next to tweets which are safe and a red indicator next to phishing tweets. The detection mechanism is designed such that it requires no extra clicks or key press. The extension works for any tweet which appears either in a user's timeline, Twitter search results or tweets on the homepage of other Twitter users. PhishAri browser extension also works for Direct Messages (DM) of a user if the URL in the DM has been detected as phishing by a blacklist. 
%In future, when we extend the scope of PhishAri and implement oauth authentication of the users, we will be able to detect phishing DMs more accurately. 
Figure~\ref{fig:extension} shows the red and green indicators at the end of the URL in each of the tweets. 

The current version of PhishAri extension works for `Chrome' browser and is written in Javascript. The browser extension extracts the tweet ID~\footnote{tweet ID is the numeric unique identifier of a tweet} of a tweet and then makes a request to the PhishAri API hosted on a separate server. The API takes the tweet ID as input and returns back a string indicating whether the tweet is `phishing' or `safe.' Accordingly, PhishAri extension displays either a red or a green indicator in front of the tweet. This whole process is very robust and it takes a maximum of 0.522 seconds for an indicator to appear for a tweet. However, this time is dependent on various factors such as the speed of feature extraction, Internet bandwidth and time to query Twitter API. We elaborate our system configuration which affects the feature extraction and classification time. Figure~\ref{fig:extension} shows a screenshot of the extension which is available on Chrome Web Store for free download.

%The indicators are loaded automatically as soon as a tweet appears once the extension is installed.

\subsection{PhishAri API}
%The decision whether a tweet is phishing or not is taken by this API. 

PhishAri API is a RESTful API written in Python using mod\_wsgi~\footnote{\url{http://code.google.com/p/modwsgi/}} framework. mod\_wsgi framework enables the Apache server to host a Python application. The API is hosted on an Intel Xeon 16 core Ubuntu server with 2.67 GHz processor and 32 GB RAM. 

The API provides a POST method to submit tweets for analysis. Once a tweet is submitted to the API, it classifies the URL as `phishing' or `safe' with the help of the set of features described in Section~\ref{section:features} using a trained classifier model pre-loaded on the server. Since our goal is to provide realtime indication to Twitter user, we require the time period for feature extraction and classification to be very less. To facilitate this, the API has multiprocessing modules which extract independent features simultaneously, hence saving a large amount of time in processing. Once the classification is done, the decision is output in form of a JSON string.  

%{\bf PK:  plan to explain this figure more somewhere? AA: Explained Below} 

\begin{figure}[!h]
\centering
\includegraphics[scale=0.26]{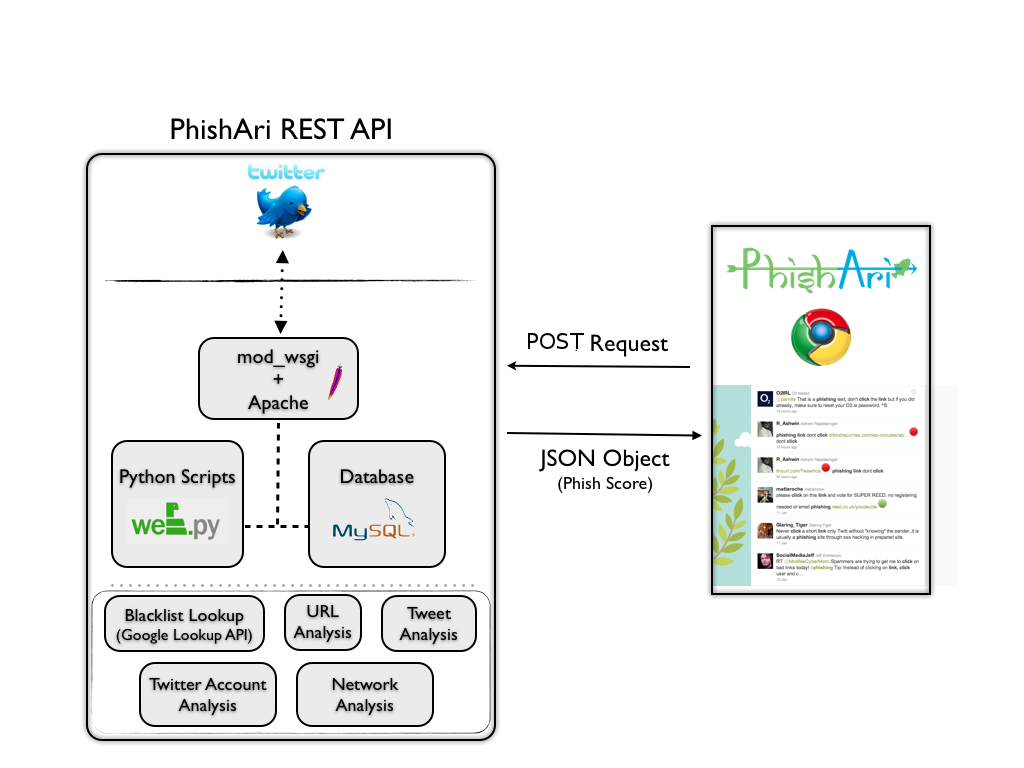}
\caption{Integration of PhishAri API with the browser extension. The extension sends tweet ids and URLs to the API through a POST request. The API responds with the results based on which the red or green indicators are embedded to the corresponding tweets by the extension}
\label{fig:arch}
\end{figure}

Figure~\ref{fig:arch} shows the integration of PhishAri browser extension with the PhishAri API. The extension sends a POST request to the API with the tweet ID. Once the API gets the tweet ID, it extracts all information about the tweet using the features mentioned in Section~\ref{section:features}. These features include URL specific features, Twitter user information and details about the Twitter network of the user. Using these features, the API constructs a feature vector which is used for classification by comparing the feature vector to a pre-loaded classifier model for phishing tweet detection. Once the decision is made, the API returns back a JSON object indicating whether the tweet is phishing or not. 

%{\bf PK:  Why do you need ``PhishAri" on top of the figure in Fig 5??? Get rid of it.... Add one or two more lines in the caption mentioning what is happening in the interaction. AR: got rid of ``PhishAri" also edited the pic to change request from GET to POST added some description}
\section{Experimental Setup}\label{section:experiment}

In this section, we describe the mechanism used for classification of phishing tweets. Our aim is to detect phishing tweets in realtime. In order to build such a mechanism, we need to identify the correct and most efficient classification methodology for which we setup the experiment. In this section, we explain the experimental setup for our study. We describe various machine learning techniques we use for phishing tweets classification. Machine learning techniques involve  classification of an unseen data point using a classification model built on a pre-labeled (already classified) dataset. Hence, our experiment involves three stages. In the first stage, to create a labelled dataset, we collect tweets with URLs and label these tweets as `phishing' or `safe.' In the second stage, we train a classifier model using a classification algorithm. In the third stage, whenever we obtain a tweet with URL, we use the trained model to make a classification decision for this newly appeared URL. 
%{\bf PK: Why explain these simple ones here, just giving the reference not enough? why waste space??? AA: we have 12 pages to fill! will remove if reqd} 
Now, we describe the machine learning algorithms we use for our study and the evaluation metrics which indicate the quality and the accuracy of our classification task.

\subsection{Machine Learning Classification}
To evaluate the most effective technique for phishing detection on Twitter, we investigate the use of multiple classification algorithms. This section explains these algorithms in brief and details on how we use them for our phishing detection task.
\subsubsection{Naive Bayes}
This is a probabilistic classifier and is based on the Naive Bayes' theorem. It works efficiently when the dimensionality of the input feature vector is high and each feature is independent of each other. Based on each feature, the Naive Bayes classifier computes the likelihood of the data point to be classified into each possible category. The data point is then classified into the category for which the likelihood (probability) is the highest. For this study, we use the \emph{naivebayes} module of Python NLTK \emph{classify} package.~\footnote{\url{http://nltk.org/api/nltk.classify.html#module-nltk.classify.naivebayes}}  
 
\subsubsection{Decision Trees}
This is a widely used machine learning technique. It is based on a predictive model which creates a classification tree. Decision tree algorithm creates a model that predicts the category of the target data point by learning simple decision rules inferred from the data features. We use `\emph{DecisionTreeClassifier}' module provided by `\emph{scikit}' library.~\footnote{\url{http://scikit-learn.org/}}
\subsubsection{Random Forest}
Random Forest is one of the most accurate classifiers and it works efficiently for large databases. For each data point to be classified, this technique randomly chooses a subset of features which are used for classification. It selects the most important features of the data point hence improves the predictive accuracy and controls over-fitting. We use `\emph{RandomForestClassifier}' module provided by  `\emph{scikit}' library for this study.

\subsection{Training and Testing Data}
We perform a 5 fold cross-validation for computing the classification results. The labeled dataset is partitioned into 5 subsets. In each test run, 4 subsets are used for training and the remaining subset is used as test data. Hence, we classify using 5 test run which ensures that each set has been used for training as well as testing. The final classification result is the average of results from the 5 classification runs.

\section{Results}\label{section:results}
As stated earlier, our study consists of two parts. In the first part, we develop a classification model based on various features like URL based and Twitter based features and classify tweets accordingly as phishing or safe. This forms our PhishAri API which uses a trained model and classifies incoming tweets based on the described features. In the next step, we create an end-user solution by deploying a Chrome extension which makes a call to the above API and public blacklists and then marks each tweet as phishing or safe with the help of a color-coded marker. 

In this section, we elaborate the results of the first part of our study, i.e., the results and observations based on the classification mechanism using the four set of feature sets described in Section~\ref{section:features}.  
\subsection{Evaluation Metrics}
In order to evaluate the effectiveness of our classification method based on the features described, we use the standard information retrieval metrics viz. accuracy, precision and recall. Precision of a class is the proportion of predicted positives in that class that are actually positive. Recall of a class is the proportion of the actual positives in that class which are predicted positive. To explain this further, we use the `confusion matrix' described in Table~\ref{tab:confusion}. 

\begin{table}[htbp]
  \centering
  \caption{Confusion matrix for classification.}
    \begin{tabular}{c|c|cc} 
    \multicolumn{2}{c}{\multirow{2}[4]{*}{}} & \multicolumn{2}{c}{\textbf{Predicted}}  \\ \hline 
    \multicolumn{2}{c}{} & \textbf{Phishing} & \textbf{Safe} \\ \hline
   \multirow{2}[4]{*}{\textbf{Actual }} & \textbf{Phishing} & TP    & FN \\  
           ~ & \textbf{Safe}  & FP    & TN  \\ \hline
    \end{tabular}%
  \label{tab:confusion}%
\end{table}

Each entry in the table indicates the number of elements of a class and how they were classified by our classification method. For example, `TP' is the number of phishing tweets which were correctly classified as phishing. Using this confusion matrix, we can compute the precision (Equation~\ref{eq1}) and recall (Equation~\ref{eq2})  for both `phishing' and `safe' classes. We also use the confusion matrix to compute the overall `accuracy' (Equation~\ref{eq3})  of the classifier. It is the ratio of the correctly classified elements of either class to the total number of elements. 

\begin{equation}\label{eq1}
Precision_{phishing} = TP / (TP + FP) 
\end{equation}
\begin{equation}\label{eq2}
Recall_{phishing} = TP / (TP + FN)
\end{equation}
\begin{equation}\label{eq3}
Accuracy = (TP + TN) / (TP + FP + TN + FN)
\end{equation}

\subsection{Classification Results}
We now describe the results of our classification experiment as described in Section~\ref{section:experiment}. We use three classification methods for our study viz. Naive Bayes, Decision Trees and Random Forest. We present the results of classification task using all these methods.

From the 1,589 phishing tweets, we found that 1,473 tweets had unique text. Therefore, is our true positive dataset, we consider these 1,473 phishing tweets and 1,500 safe tweets chosen randomly from the tweets marked as `safe' during our data collection process. We use this dataset for the rest of our classification experiments. Previous studies show that there is 8\% spam content on Twitter which consists of phishing, malware and other unwanted tweets~\cite{grier2010spam}. Therefore, to balance the prediction error and minimize the overall error rate, we assign positive weights to spam class to account for the unbalanced dataset. We found that Random Forest classifier works best for phishing tweet detection on our dataset with a high accuracy of 92.52\%. We also obtain a recall of 92.21\% for phishing class and 96.82\% for safe class. The results from the three classification techniques are described in the table~\ref{tab:results}. It is observed that when we used Random Forest classifier, we also achieved a high recall and precision for both `phishing' and `safe' classes. It is important in our study to achieve a good precision of both classes to reduce the number of false negatives and false positives. Precision-accuracy balance is hard to achieve and we notice that the precision of phishing class drops but accuracy increases when we move from Naive Bayes classifier to Decision Tree classifier. However, we finally achieved a desirable precision and accuracy when we used Random Forest classifier. Random Forest reduces false positives and hence the precision of both the classes increased significantly.

\begin{table}[htbp]
\centering
\caption{Results of classification experiments. We observe that Random Forest performs the best with an accuracy of 92.52\%}
\begin{tabular}{|p{2.3cm}|p{1.5cm}|c|p{1.1cm}|} 
      \hline
\textbf{Evaluation metric} & \textbf{Naive Bayes} & \textbf{Decision Tree} & \textbf{Random Forest} \\ \hline
\textbf{Accuracy} & 87.02\% & 89.28\% & 92.52\%  \\ 
\textbf{Precision (phishing)} & 89.21\% & 88.05\% & 95.24\% \\ 
\textbf{Precision (safe)} & 92.12\% & 94.15\% & 97.23\%  \\ 
\textbf{Recall (phishing)} & 68.32\% & 74.51\% & 92.21\% \\ 
\textbf{Recall (safe)} & 85.67\% & 89.20\% & 95.54\% \\ \hline
\end{tabular}
\label{tab:results}
\end{table}

Previous studies show that Random Forest outperforms all classifiers for phishing email detection with an error rate of 7.72\%~\cite{abu2007comparison}. We find that the superior performance of Random Forest for phishing detection on Twitter also holds true with a high accuracy. 
We further investigate the performance of Random Forest classification method by using the confusion matrix described in Table~\ref{tab:confusion_rand}. We show that we could detect 92.31\% phishing tweets correctly. However, we misclassified 9.6\% of legitimate tweets as phishing tweets. This is because the user behaviour of the many of the source Twitter users of such tweets is very close to that of a phisher like - extensive use of unrelated hashtags and automated tweet activity. The false negative percentage is low indicating that we classified only 7.78\% phishing tweets as legitimate. The misclassification of phishing tweets as legitimate tweets happens because some phishing tweets exhibit similar features as legitimate tweets. We manually observed a sample of such misclassified tweets and found that there are Twitter accounts which often exhibit dual behavior by sometimes posting legitimate tweets and sometimes phishing tweets. These users are either already compromised or due to negligence, retweet a phishing tweet. Hence, tweets from such users are misclassified,  as their behavior and attributes are very similar to both legitimate users and phishers. Since our classification methodology takes into account Twitter based features, with the evolution of phishing techniques on Twitter, if a malicious user makes the phishing tweet look like a legitimate tweet and has Twitter network features as that of a legitimate user, our classification method may misjudge the phishing tweet as legitimate. 
\begin{table}[htbp]
  \centering
  \caption{Precision and Recall for phishing detection using Random Forest based on all four feature sets.}
    \begin{tabular}{c|c|cc} 
    \multicolumn{2}{c}{\multirow{2}[4]{*}{}} & \multicolumn{2}{c}{\textbf{Predicted}}  \\ \hline 
    \multicolumn{2}{c}{} & \textbf{Phishing} & \textbf{Safe} \\ \hline
   \multirow{2}[4]{*}{\textbf{Actual} } & \textbf{Phishing} & \textbf{92.31\%}    & 7.78\% \\  
           ~ & \textbf{Safe}  & 9.60\%    & \textbf{94.41\%}  \\ \hline
    \end{tabular}%
  \label{tab:confusion_rand}%
\end{table}

\subsection{Evaluation of various Feature Sets}
Most of the previous studies to detect phishing have used features based on the URL of the suspicious page and the HTML source of the landing page. In this study, we propose to use Twitter based features along with URL based features to quickly detect phishing on Twitter at zero-hour. To evaluate the performance of detection using these additional set of features based on Twitter properties, we present feature-set wise performance of the classification technique we use. 

\begin{table*}[!htbp]
\centering
\caption{Feature set wise performance of classification of Phishing Tweets.}
\begin{tabular}{|l|c|c|c|c|c|} \hline
\textbf{Feature Sets} & \textbf{Precision (Phishing)} & \textbf{Precision (Safe)} & \textbf{Recall (Phishing)} & \textbf{Recall (Safe)} & \textbf{Accuracy} \\ \hline
\textbf{F1} & 81.27\% & 88.21\% & 79.25\% & 91.34\% & 82.22\% \\ 
\textbf{F1 + F2} & 86.11\% & 89.92\% & 85.21\% & 92.21\% & 87.31\% \\ 
\textbf{F1 + F2 + F3} & 91.10\% & 94.66\% & 88.32\% & 92.88\% & 90.03\% \\ 
\textbf{F1 + F2 + F3 + F4} & 95.24\% & 97.23\% & 92.21\% & 95.54\% & 92.52\%  \\ 
\hline \end{tabular}
\label{tab:setwise_eval}
\end{table*}

As described in Table~\ref{tab:features_table}, we have used four sets of features in this study. To evaluate the impact of each feature set, we performed classification task by taking one feature set at a time and then added the other one in the next iteration. Table~\ref{tab:setwise_eval} presents our experiment results by using different set of features using Random Forest classification method which gives us the overall highest accuracy of 92.52\%. We observe that when we use only URL based features, we get an overall accuracy of 82.22\% and a low precision and recall for `phishing' class. The addition of Twitter based feature sets, user based features and network based features significantly improve the performance of phishing detection and boost the precision of identifying phishing tweets significantly. Hence, Twitter based features are helpful in increasing the performance of classifying phishing tweets.

\subsection{Most Informative Features}
We now evaluate the most important features which help to decide whether a tweet is phishing or not. We use `scikit' library to find out the most informative features. Random Forests deploy ensemble learning to evaluate the feature importance. After each random tree is constructed using a set of features, its performance (misclassification rate) is calculated. Then the values of each feature is randomly permuted (for each feature) and the new misclassification rate is evaluated. The best performing features are then chosen as the most informative features. The most informative features which we found for phishing tweet detection using Random Forest classification are described in Table~\ref{tab:important_feats}. 

Ownership period is one of the most important features in phishing detection. The domains of malicious and phishing URLs tend to be short lived when compared to the domains of legitimate URLs in order to avoid detection. Similarly the age of Twitter account of the user posting phishing tweets is also generally less. Such users are often detected by Twitter and their accounts are suspended. However, using PhishAri API, we could detect a large number of phishing tweets by such users before they were suspended by Twitter.

Another important feature is the presence of conditional redirects. Many phishing websites redirect the user to a legitimate website instead of the phishing landing page if the page is being accessed via an automated script or bot. In our experiment, we compare the landing URL when the suspected URL is accessed by the browser simulation and bot simulation. In case the landing URLs are different, there is a high possibility that the website is malicious. The redirection to a legitimate website when accessed by an automated script is to avoid detection by bots such as googlebots traversing through the Internet.

We also find that presence of trending \#tags in a tweet is an important feature for phishing detection. Phishers often hijack trending topics and start posting unrelated content in their tweets with the trending \#tag appended. This increases the visibility of their tweet as trending topics specific to a location are always displayed on the homepage of a Twitter user.

Phishers usually have more number of followees than followers. Since relationships on Twitter are unidirectional, a Twitter user needs trust to be followed by another user. Since phishers do not often post legitimate content, very few Twitter users tend to follow phishers. However, phishers follow a lot of users in the hope of being followed back. Hence the ratio of Follower-Followee is very skewed in case of phishing tweets.

Another technique used by phishers to gain visibility is to directly mention other Twitter users in their tweets. Phishers tend to have a lot of @tags in their tweets so that their tweet is directly visible to the mentioned users. Since the mentioned users receive these tweets in their timeline, there is a high chance that the target users click on the links and fall victim to phishing attacks. 

\begin{table}[!htbp]
\centering
\caption{Most informative features for detecting phishing tweets.}
    \begin{tabular}{|c|l|}
        \hline
        \textbf{Ranking} & \textbf{Feature}                           \\ \hline
        1       & Ownership period                  \\ 
        2       & Age of account                    \\ 
        3       & Presence of conditional redirects \\ 
        4       & Presence of trending \#tags        \\ 
        5       & Number of Redirections            \\ 
        6       & Follower-Followee Ratio           \\ 
        7       & Number of @tags                   \\
        \hline
    \end{tabular}
\label{tab:important_feats}
\end{table}

\subsection{Comparison of PhishAri with Blacklists}
The inherent problem of the blacklists is that they are slow to catch phishing URLs. Since Twitter provides a realtime stream of tweets to a user, it is important that the tweets are detected as phishing as soon as they appear to the user. Blacklists in such cases prove to be ineffective. To support our claim, we compare the performance of PhishAri with two public blacklists, Google Safebrowsing and PhishTank.

At the time of data collection, we collected realtime stream of tweets from Twitter and immediately look up the URLs present in the tweet in these blacklists. Since blacklists take some time to add newly created phishing URLs, we wait for 3 days and again lookup the URLs collected 3 days ago in the Google safebrowsing and PhishTank blacklists. We also use PhishAri to classify each of these tweets as phishing or safe. 

We found that 80.6\% unique phishing tweets were detected as phishing at zero-hour by PhishAri which were caught by the blacklists only later when we checked after 3 days.  Public blacklists are often based on crowdsourcing (like PhishTank) or use URL based or landing page based features. However, phishers often keep changing their strategies and hence these detection mechanisms by blacklists often fail. We couple these features along with other features for a better phishing detection to obtain efficient realtime detection. This shows that PhishAri can complement the blacklisting mechanism for Twitter to detect more phishing URLs in realtime.

\subsection{Comparison of PhishAri with Twitter}
Twitter has its own detection mechanism for catching malicious, spam and phishing tweets. In case a URL in a tweet is not safe, Twitter shows a warning page to the user when one tries to navigate to that URL from Twitter. However, we found that Twitter's mechanism was not as quick and was unable to catch a large fraction of phishing URLs appearing in tweets in realtime. 

%{\bf PK:  Can you add the number of URLS that you tested in a line here} 
To compare the performance of PhishAri API with Twitter's detection mechanism, we check whether Twitter marks a URL as safe or not at the time it is submitted to Twitter stream. Then, we again check the status of the URL after 3 days. Out of 3,09,321 tweets with URLs, we found that 492 tweets were undetected by Twitter at the time of data collection, however they were marked as `suspicious' URLs only later when we checked after 3 days. However, PhishAri was able to detect 84.6\% of these phishing tweets at zero-hour which were blacklisted by Twitter later. This shows that PhishAri if implemented along with Twitter's malicious tweets detection mechanism, can help boost the performance of realtime detection of phishing on Twitter.

\begin{figure*}[!htbp]
\centering
\mbox{
\subfigure[]{
\includegraphics[width=.4\textwidth]{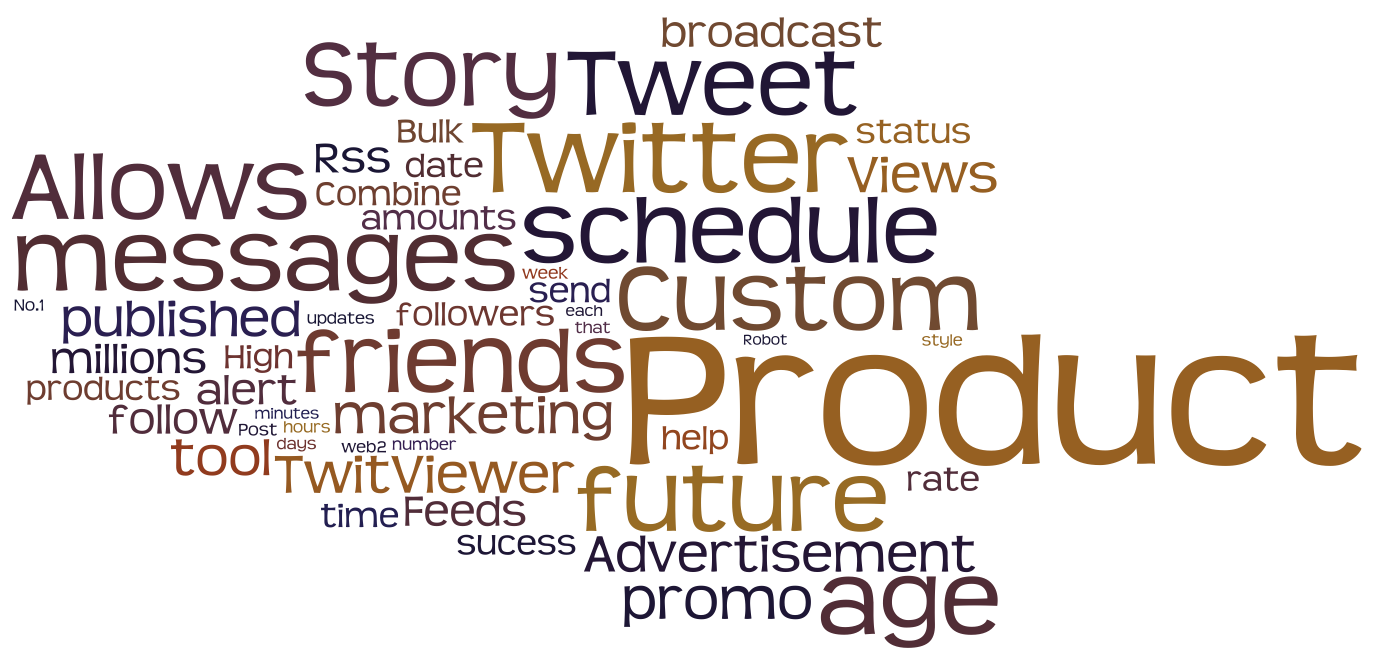}
\label{fig:tagcloud}
}\quad
\subfigure[]{
\includegraphics[width=.4\textwidth]{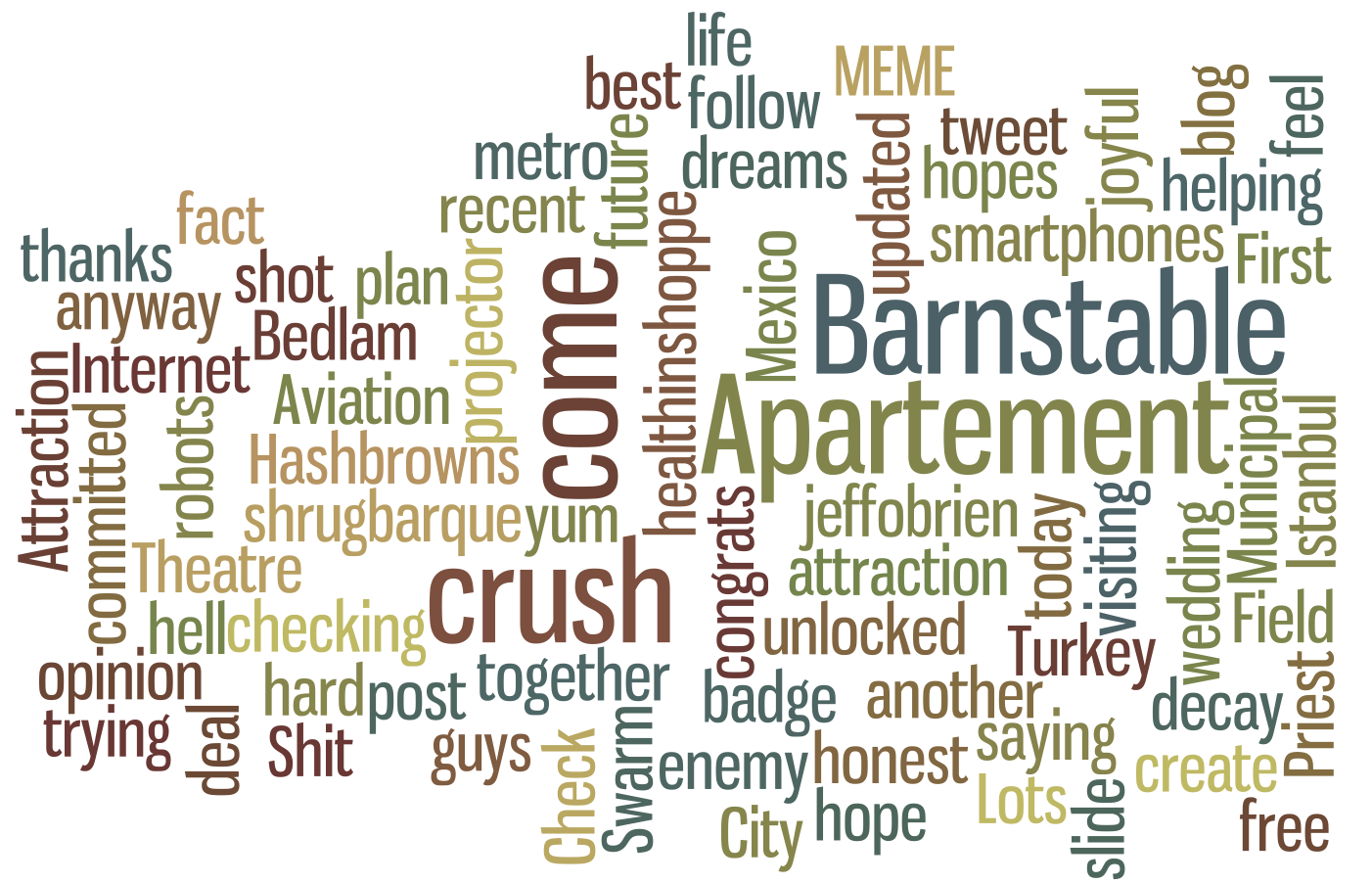}
\label{fig:legit_tagcloud}
}
}
\centering
\caption{Figure~\ref{fig:tagcloud} shows the most frequent words of phishing tweets in our dataset. Figure~\ref{fig:legit_tagcloud} shows the frequent words occurring in a sample of legitimate tweets from our dataset. Both the tagclouds have random 50 tweets. In case of phishing tweets there is a dominance of certain words which signify the spam campaign promoted at that time, however, the legitimate tweets have almost all words occurring with equal probability.}
\label{fig-all}
\end{figure*}

\subsection{Time Evaluation}
One of the major aims of this study is realtime detection of phishing tweets. Hence our mechanism needs to be robust enough to quickly classify a phishing tweet. We now evaluate how much time PhishAri takes to classify a URL. As mentioned before, a classifier model is preloaded on our server which is used to make decisions about a tweet. The PhishAri API is written using multiprocessing modules so that it can extract independent features simultaneously, hence increasing the speed of computation. We find that the time required for the feature extraction and classification of a tweet is a maximum of 0.522 seconds (Min: 0.167 sec, Avg: 0.425 sec, Median 0.384 sec). This time was taken when we ran our experiments on an Intel Xeon 16 core Ubuntu server with 2.67 GHz processor and 32 GB RAM. However, we must note that the speed of classification is also dependent on the response times of the Twitter API, the WHOIS repository and also the Internet bandwidth.

\subsection{Characteristics of Phishing Tweets}
We also found that the words used in case of phishing tweets are different from those used in legitimate tweets. Phishing tweets often have keywords which are specifically used to lure the unsuspecting Twitter user into clicking the URL. The content of the tweet is often appealing enough and promises some kind of benefit to the user if one visits the URL. Figure~\ref{fig:tagcloud} shows the most popular words which appear in phishing tweets. We see that `product,' `allow,' etc. are the most popular words. They appear repeatedly because of a phishing campaign which asks Twitter details in return for more Twitter followers. 

The text of phishing tweets is considerably different from that of legitimate tweets, where people usually talk about general topics and use a variety of words unlike phishing tweets which use a limited set of words. Figure~\ref{fig:legit_tagcloud} shows the word tag cloud of a sample of legitimate tweets. The words occurring in legitimate may also depend on the trending hashtags at the time tweets were posted. However, the text for phishing tweets remains relatively the same for a particular phishing campaign irrespective of the trending topic. However, phishing tweets contain the hastags which are trending at the time they were posted to gain visibility.

\begin{figure}[!ht]
\centering
\includegraphics[scale=0.45]{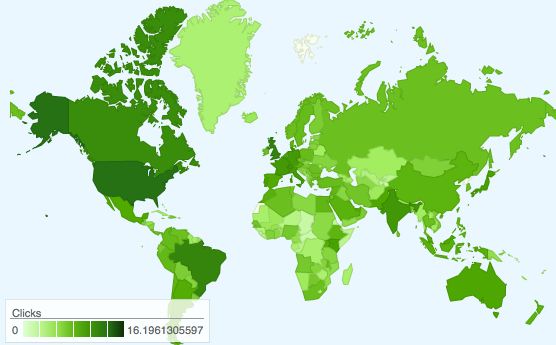}
\caption{Countries from where phishing tweets originate.}
\label{fig:geomap}
\end{figure}

We also try to ascertain the country of the origin of phishing tweets in our dataset. We find that USA has maximum number of users posting phishing URLs followed by Brazil. The geomap in Figure~\ref{fig:geomap} shows the concentration of phishing URLs originating from various countries across the world on Twitter. Manual evaluation shows that many of the phishing accounts were indeed from USA. However, it must be noted that the phishers could've falsely selected the country as USA in their Twitter bio page. Also, more than 25\% of all Twitter users are from USA, thus, it might seem natural that there are more phishing tweets originating from there.~\footnote{\url{http://venturebeat.com/2012/07/30/twitter-reaches-500-million-users-140-million-in-the-u-s/}}

\section{PhishAri Extension for Chrome Browser}\label{section:phishari_extension}
In this section, we evaluate the realtime browser extension we built for phishing detection. The extension works for Chrome browser and has currently more than 70 active users. We evaluate user experience of our extension and present statistics about how our extension is being used by Twitter users. 
\subsection{User Experience}
%{\bf PK:  put some references like the privacy bird user study here so reviewers know this is a standard methodology and say, researchers have done such evaluations before} 
There have been user studies to evaluate and assess the user-experience of browser based tools~\cite{cranor2002use}. The users of the study are asked to use the tool and give feedback about the system. We performed a lab study to find out the user experience with PhishAri and whether users find the extension effective and useful. The lab study consisted of 10 users out of which 7 were males and 3 females. All users were active on Twitter. They were given the download link of the PhishAri extension page which also gave details about the extension and described how it worked. 
%Then after installation of the extension, they were asked to browse Twitter and give feedback about the extension.
Each user was asked to browse Twitter after installing the PhisAri extension and were asked if they found the plugin effective and easy to use. All users said that the extension was very easy to use and the indicator displayed with every tweet did not adversely affect their Twitter experience. However, 4 users commented that they prefer using Twitter clients over browser. Currently, PhishAri provides support only for browser based Twitter access. For all 10 users, the color coded indicators appeared as soon as the tweet loaded without any visible delay. However, 5 users observed a time lag in the appearance of the indicators when they were browsing tweet stream of a trending topic on Twitter. In future, we will try to make PhishAri faster to fill this gap. Since phishing tweets are not abundantly present as compared to other tweets, and we conducted a time-limited lab study, we created a dummy Twitter account which had a mix of phishing and legitimate tweets. Users were asked to go to the dummy Twitter account to check if a red indicator appears for phishing tweets or not. Two users commented that whenever there is a red indicator for a tweet, they would like to see a preview of the landing page (as in web-based systems likes \emph{PhishTank}) when they hover over the indicator. We think that adding this feature in future would be useful to gain the confidence of user.

Users were asked if they would be interested to use the PhishAri extension daily in regular use. Except the 4 users who prefer using Twitter client over browser, all other users said that PhishAri seems to be a useful tool. These users also said that they would like to have a similar spam detection tool for Twitter which indicates whether a tweet (irrespective of the presence of a URL) is spam or legitimate. However, the scope of PhishAri is currently to indicate whether a tweet which has a URL is `phishing' or `safe.' The lab study showed that PhishAri works with ease and is non-intrusive; the indicators do not distract the users attention while browsing Twitter. The color coded indicators are effective in indicating the status of a tweet but could be improved by showing an optional preview of the landing page when the cursor is hovered over the indicator. 

\subsection{Statistics}
We present some statistics about PhishAri browser extension. We have Google Analytics~\footnote{\url{http://www.google.com/analytics}} enabled for our extension which helps us track the user details like the country and the active time of user using the extension. We found that we have a wide diversity of users from various  countries with highest traffic from the US and India. Table~\ref{tab:user_countries} shows the percentage of users from various countries who use PhishAri. 
% {\bf PK:  surprising to have US here??? we are propagating there right?} AA: Not really, not sure how come US visitors are so high. Maybe through our FB network

\begin{table}[htbp]
\centering
\caption{PhishAri Chrome extension users from various countries across the world.}
����\begin{tabular}{|l|l|} \hline
��������\textbf{Country / Territory} & \textbf{Users}�� \\ \hline
��������United States������ & 32.59\% \\ 
��������India�������������� & 28.09\% \\ 
��������Germany������������ & 8.20\%� \\ 
��������Saudi Arabia������� & 6.90\%� \\ 
��������United Kingdom����� & 3.62\%� \\ 
��������Greece������������� & 3.35\%� \\ 
��������France������������� & 2.93\%� \\ 
��������Russia������������� & 2.70\%� \\ 
��������Slovakia����������� & 2.25\%� \\ 
��������Egypt�������������� & 2.09\%� \\ 
��������Singapore���������� & 1.41\%� \\ 
��������Morocco������������ & 1.29\%� \\
	\hline \end{tabular}
\label{tab:user_countries}
\end{table}

\section{Discussion}\label{section:discussion}
In this section we highlight some important aspects of PhishAri. 
\paragraph{Selection of features for realtime detection}
There have been studies which show that extraction of Twitter user specific information helps in a very accurate spam detection. Since we wanted our system to be fast, efficient and executable in realtime, we experimented and discarded features which included analyzing all tweets by the source user in favor of faster system performance. We observed that Twitter user specific features like comparison of text of all the user's tweets and finding features related to the Twitter friends of that user do not increase the classification accuracy but significantly increase the response time. Hence, we discard such features and yet achieve an accuracy of 92.52\%. Carefully chosen important features based on URL analysis, tweet analysis and tweet user's analysis, help us to detect phishing with high confidence but in a reasonable amount of time so that end users can use our methodology in practice. 

\paragraph{Parallel computation of features} 
To enable quick decision on a tweet, we have multiprocessing modules in our system which extract features in parallel. This helps in reduction of overall computation time. However, in future, we can further improve the feature extraction by distributing the computation of features across multiple servers.

\paragraph{PhishAri available as API}
PhishAri is available as a RESTful API which can be called using an HTTP POST request by passing the tweet ID as the input parameter. We have yet not (but soon to be) released the PhishAri API publicly, but it can be used by various applications to decide whether a tweet is phishing or not. It can also be a complementary technique used along with Twitter's defense mechanism for better protection from phishing on Twitter.

\section{Future Work}\label{section:future}
Now we discuss how we can further improve PhishAri for more efficient and robust phishing detection.

\paragraph{Backend database for faster lookup}
In future, we can maintain a cache backend database to capture tweets which have already been marked as either phishing or safe on Twitter. So, if a tweet with same URL appears on Twitter, then we can skip the entire process of feature extraction and classification and lookup in our dataset of phishing URLs and safe URLs. This will also help us increase our own database of phishing tweets.

\paragraph{Increase the scope of PhishAri from public to all tweets}
Currently, PhishAri can detect whether a tweet is phishing or not only if the source Twitter user of that tweet is a public user. Otherwise PhishAri is unable to extract the user specific information. In future, we will implement oauth integration of Twitter with PhishAri so that it can detect a wider range of phishing tweets. However, this is just a proof of concept and does not affect our methodology in any way.

\section{conclusion}\label{section:conclusion}
In this study, we built PhishAri -- an effective mechanism to detect phishing on Twitter. Our methodology exploits not just the traditional phishing detection features which are based on the URL and the suspicious landing page, but also Twitter specific and WHOIS based features. We use a combination of URL based and Twitter based features which help in an effective and realtime detection of phishing on Twitter. As a proof of concept, we also develop a RESTful API which can be accessed using an HTTP POST method. We also implement a Chrome browser extension which makes a call to this API and accordingly shows an indicator next to each tweet indicating whether the tweet is phishing or not. We also show that our methodology works faster than standard blacklisting mechanism and Twitter's own defense mechanism. We were able to detect 80.6\% more URLs than popular blacklists like PhishTank and Google Safebrowsing at zero-hour with an accuracy of 92.52\%. Similarly, our detection mechanism also works better than Twitter's defense system by 84.6\% at zero-hour. Since we do not achieve a 100\% accuracy, there is always a possibility of false negatives. However, our method can be coupled with blacklisting and Twitter's defense mechanism for a better, more accurate realtime detection of phishing on Twitter.

\section*{Acknowledgments}

We thank all members of PreCog research group at IIIT-Delhi for their valuable feedback and suggestions. Authors would also like to thank Aditi Gupta for her feedback on initial drafts of this paper. This work was done when Ashwin was interning at PreCog. 

%PK: Add this only if you think she has done at least a couple of thorough reviews.... You can give her one now to review, if you want.... Authors thank Aditi Gupta for her feedback on initial drafts of this paper.

%
%{\bf PK:  put the same text as in CEAS appropriately...The authors would like to thank...}



%\bibliographystyle{IEEEtran}

\bibliographystyle{IEEEtran} 
\bibliography{phishariref}

\end{document}